\documentstyle[art12,epsf]{article}
\newcommand{\be}{\begin{equation}}
\newcommand{\ee}{\end{equation}}
\newcommand{\bea}{\begin{eqnarray}}
\newcommand{\eea}{\end{eqnarray}}

\newcommand{\p}{\partial}
\newcommand{\s}{\sigma}

\newcommand{\up}{\uparrow}
\newcommand{\down}{\downarrow}
\newcommand{\la}{\langle}
\newcommand{\ra}{\rangle}
\newcommand{\rd}{\mbox{d}}
\newcommand{\ri}{\mbox{i}}
\newcommand{\re}{\mbox{e}}

\def\3he{{$^3${\rm He}}}

\def\slD{\raise.15ex\hbox{$/$}\kern-.57em\hbox{$D$}}
\def\dsl{\raise.15ex\hbox{$/$}\kern-.57em\hbox{$\Delta$}}
\def\slp{{\raise.15ex\hbox{$/$}\kern-.57em\hbox{$\partial$}}}
\def\nsl{\raise.15ex\hbox{$/$}\kern-.57em\hbox{$\nabla$}}
\def\sla{\raise.15ex\hbox{$/$}\kern-.57em\hbox{$\rightarrow$}}
\def\slla{\raise.15ex\hbox{$/$}\kern-.57em\hbox{$\lambda$}}
\def\gtwid{\raise.3ex\hbox{$>$\kern-.75em\lower1ex\hbox{$\sim$}}}
\def\ltwid{\raise.3ex\hbox{$<$\kern-.75em\lower1ex\hbox{$\sim$}}}

\def\12{{1\over2}}

\def\part{\partial}
\def\la{\lambda}

\def\be{{\beta}}

\def\bethlogo{\vbox{\bf \line{\hrulefill} 
    \kern-.5\baselineskip 
    \line{\hrulefill\phantom{ ELIZABETH A. MASON }\hrulefill} 
    \kern-.5\baselineskip 
    \line{\hrulefill\hbox{ ELIZABETH A. MASON }\hrulefill} 
    \kern-.5\baselineskip 
    \line{\hrulefill\phantom{ 1411 Chino Street }\hrulefill} 
    \kern-.5\baselineskip 
    \line{\hrulefill\hbox{ 1411 Chino Street }\hrulefill} 
    \kern-.5\baselineskip 
    \line{\hrulefill\phantom{ Santa Barbara, CA 93101 }\hrulefill} 
    \kern-.5\baselineskip 
    \line{\hrulefill\hbox{ Santa Barbara, CA 93101 }\hrulefill}
    \kern-.5\baselineskip 
    \line{\hrulefill\phantom{ (805) 962-2739 }\hrulefill} 
    \kern-.5\baselineskip 
    \line{\hrulefill\hbox{ (805) 962-2739 }\hrulefill}}}
\def\lisalogo{\vbox{\bf \line{\hrulefill} 
    \kern-.5\baselineskip 
    \line{\hrulefill\phantom{ LISA R. GOODFRIEND }\hrulefill} 
    \kern-.5\baselineskip 
    \line{\hrulefill\hbox{ LISA R. GOODFRIEND }\hrulefill} 
    \kern-.5\baselineskip 
    \line{\hrulefill\phantom{ 6646 Pasado }\hrulefill} 
    \kern-.5\baselineskip 
    \line{\hrulefill\hbox{ 6646 Pasado }\hrulefill} 
    \kern-.5\baselineskip 
    \line{\hrulefill\phantom{ Santa Barbara, CA 93108 }\hrulefill} 
    \kern-.5\baselineskip 
    \line{\hrulefill\hbox{ Santa Barbara, CA 93108 }\hrulefill}
    \kern-.5\baselineskip 
    \line{\hrulefill\phantom{ (805) 962-2739 }\hrulefill} 
    \kern-.5\baselineskip 
    \line{\hrulefill\hbox{ (805) 962-2739 }\hrulefill}}}

\def\la{{\lambda}}

\def\low#1{\lower.5ex\hbox{${}_#1$}}
\def\ltwid{\raise.3ex\hbox{$<$\kern-.75em\lower1ex\hbox{$\sim$}}}

\def\psl{\raise.15ex\hbox{$/$}\kern-.57em\hbox{$\partial$}}
\def\partt{\raise.15ex\hbox{$\widetilde$}{\kern-.37em\hbox{$\partial$}}}
\def\parts{\raise.15ex\hbox{$/$}{\kern-.6em\hbox{$\partial$}}}
\def\nablas{\raise.15ex\hbox{$/$}{\kern-.6em\hbox{$\nabla$}}}
\def\oprod{\hbox{$\rm O$}{\kern -0.8em\hbox{$\Pi$}}}
\def\partw#1{\raise.15ex\hbox{$/$}{\kern-.6em\hbox{${#1}$}}}

\def\si{{\sigma}}

\def\gtappr{{{\lower4pt\hbox{$>$} } \atop \widetilde{ \ \ \ }}}
\def\ltappr{{{\lower4pt\hbox{$<$} } \atop \widetilde{ \ \ \ }}}

\def\topppageno1{\global\footline={\hfil}\global\headline
={\ifnum\pageno<\firstpageno{\hfil}\else{\hss\twelverm --\ \folio
\ --\hss}\fi}}

\def\toppageno2{\global\footline={\hfil}\global\headline
={\ifnum\pageno<\firstpageno{\hfil}\else{\rightline{\hfill\hfill
\twelverm \ \folio
\ \hss}}\fi}}

\def\ltdash{\raise-1.8pt\hbox{$\scriptscriptstyle |$}}
\def \ra{\rangle}
\def\la{\langle}

\def\s{\sigma}

\def\dg{{^
{\dag}}}

\def\ra{\rangle}
\def\la{\langle}

\def\1{{\bf 1}}
\def\2{{\bf 2}}

\def\rarrow{\rightarrow}

\def\ell{{\it l } {\rm n}}

\def\si{\sigma}

\def\cx2{\sqrt{c^2_x+c^2_y}}

\def\gkk{\gamma _{\vec k}}
\def\gk2{\gkk ^2}
\def\dw{\downarrow}
\def\up{\uparrow}
\def\gtappr{{{\lower4pt\hbox{$>$} } \atop \widetilde{ \ \ \ }}}
\def\ltappr{{{\lower4pt\hbox{$<$} } \atop \widetilde{ \ \ \ }}}

\def\pbar{{\partial\kern-1.2ex\raise0.25ex\hbox{/}}}
\def\up{\uparrow}
\def\dw{\downarrow}

\def\s{\sigma}

\def\dg{{^{\dag}}}

\def\ra{\rangle}
\def\la{\langle}

\def\1{{\bf 1}}
\def\2{{\bf 2}}

\def\rarrow{\rightarrow}

\def\ell{{\it l } {\rm n}}

\def\si{\sigma}

\def\cx2{\sqrt{c^2_x+c^2_y}}

\def\gkk{\gamma _{\vec k}}
\def\gk2{\gkk ^2}
\def\gtappr{{{\lower4pt\hbox{$>$} } \atop \widetilde{ \ \ \ }}}
\def\ltappr{{{\lower4pt\hbox{$<$} } \atop \widetilde{ \ \ \ }}}

\def\thickra{\hbox{\raise0.2pt\hbox{{$\bf >\mkern-13mu>\mkern-13mu>$}}}}
\def\thickrarrow{\hbox{\raise0.28pt\hbox{{$\bf >\mkern-13mu>\mkern-13mu>$}}}}

\begin{document}
\bibliographystyle {plain}
%\tableofcontents  
\begin{titlepage}
\begin{flushright}
\today
\end{flushright}
\vspace{0.5cm}
\begin{center}
{\Large {\bf Reflections on the One Dimensional Realization of 
Odd-Frequency Pairing}}\\
\vspace{1.8cm}
\vspace{0.5cm}
{P. Coleman$~^{1}$, A. Georges$~^{2}$ and  A. M. Tsvelik$~^{3}$}\\
\vspace{0.5cm}
{\em$~^1$ Serin Physics Laboratory, Rutgers University, P.O. Box 849, 
Piscataway, NJ}\\
{\em$^{2}$ Laboratoire de Physique Theorique de l' Ecole Normale 
Superieure,\\
24 Rue Lhomond, 75231 Paris Cedex 05-France}\\
{\em$^3$Department of Physics, University of Oxford, 1 Keble Road,}
\\
{\em Oxford, OX1 3NP, UK}\\

\begin{abstract}
\par
We discuss the odd-frequency pairing correlations
discovered by Zachar, Kivelson and Emery (ZKE) in a one
dimensional Kondo lattice. A  specific lattice model
that realizes the continuum theory of ZKE is introduced and
the correlations it gives rise to are identified
as odd-frequency singlet pairing.
The excitation spectrum is found to contain a spin gap,
and a much lower energy band of spinless excitations.
We  discuss how the power-law
correlations realized in the ZKE model evolve into true long-range
order when Kondo chains are weakly coupled together and tentatively
suggest a way in which the higher dimensional model can be treated
using mean-field theory.
\end{abstract}
%cond-mat/96
\end{center}

\end{titlepage} 
%\newpage
\sloppy
\par
\section{Introduction}

The  concept  of odd-frequency pairing as a new symmetry class
of superfluidity was conceived  twenty five years ago by 
Berezinskii \cite{ber}. 
It is well-known
that the development of a paired state
in a system with repulsive interactions is aided by the formation
of a pair-wavefunction with nodes. Berezinskii's idea
extends this concept,
proposing that superfluidity can result from a pair-wavefunction
with a node in {\sl time}.

In the years that have lapsed since Berezinskii's original
proposal, 
theoretical attempts to develop 
Berezinskii's radical concept
have been  thwarted by the absence of a weak-coupling
realization of the phenomenon.  The Landau
school of physics found early on that there were no logarithmic
singularities in the odd-frequency pairing susceptibility:
the absence of a weak coupling Cooper instability 
meant that a controled  weak-coupling treatment of the
idea was not possible. 

Five years ago, Abrahams and Balatsky\cite{bal} revived the 
idea of odd-frequency pairing, suggesting that
strong-coupling realizations of the phenomenon might be found.
They pointed out that both triplet and singlet realizations
of odd-frequency pairing are allowed by symmetry.  Efforts to
pursue this idea led to the following 
developments:
\begin{itemize}

\item{}
Emery and Kivelson \cite{emery} 
observed that odd-frequency pairing
can be regarded as the condensation of a composite order parameter.
For example, the scalar combination of a triplet pair with a spin
operator gives rise to odd-frequency singlet pairing.
The combination of a singlet pair with a spin operator gives
rise to odd-frequency triplet pairing.  

\item{} Coleman, Miranda and Tsvelik\cite{we}
have combined these ideas with the technology of Majorana fermions
to develop a mean-field treatment of odd-frequency
triplet pairing within a Kondo lattice model, suggesting 
odd-frequency triplet pairing as an alternative scenario for
heavy fermion superconductivity. In this model, is was possible to
show that a staggered composite order parameter led to a finite
Meissner stiffness. 

\item{}Balatsky, Abrahams,
Scalapino and Schrieffer \cite{abrahams}  have pursued this idea using a
composite BCS-type Hamiltonian. 
\end{itemize}
These efforts have all added plausibility 
to the Abrahams-Balatsky proposal, but the continued absence of 
a controlled, solvable model has led to a cautious 
response from the community.

Recent 
nonperturbative results due to 
Zachar, Kivelson and Emery\cite{zachar,zachar2} (ZKE)
open up an exciting new possibility. 
These authors have considered a variant of the
one dimensional Kondo lattice, and by the application of 
bosonization techniques have shown there are strong odd-frequency
pair correlations in this model.  The ZKE results strongly
suggest that a higher dimensional version of their model
would develop long-range odd-frequency pairing. 
In this paper 
we explore consequences of this non-perturbative solution.
We introduce a lattice
model where the absence of backscattering removes some  uncertainties
present in the original work. The power-law
correlations are identified as
odd-frequency singlet pairing; we  discuss how they
evolve into a state of true long-range order
when Kondo chains are weakly coupled together. 

\section{Kondo chain without backscattering}

The model suggested by ZKE for realization of odd-frequency 
pairing is a one-dimensional Kondo lattice model.
A critical and subtle  point in their arguments, was the assumption that
back-scattering off the local moments can be neglected.
To begin our discussion of their results, we shall introduce
a lattice variant of their one-dimensional model, where
back-scattering is either absent, or strongly suppressed.  

\begin{figure}[tb]
% ********   This is for two columns
%\epsfxsize=3.5in 
% ***********For one column  ********************
\epsfxsize=4.5in 
% ***********************************8
\centerline{\epsfbox{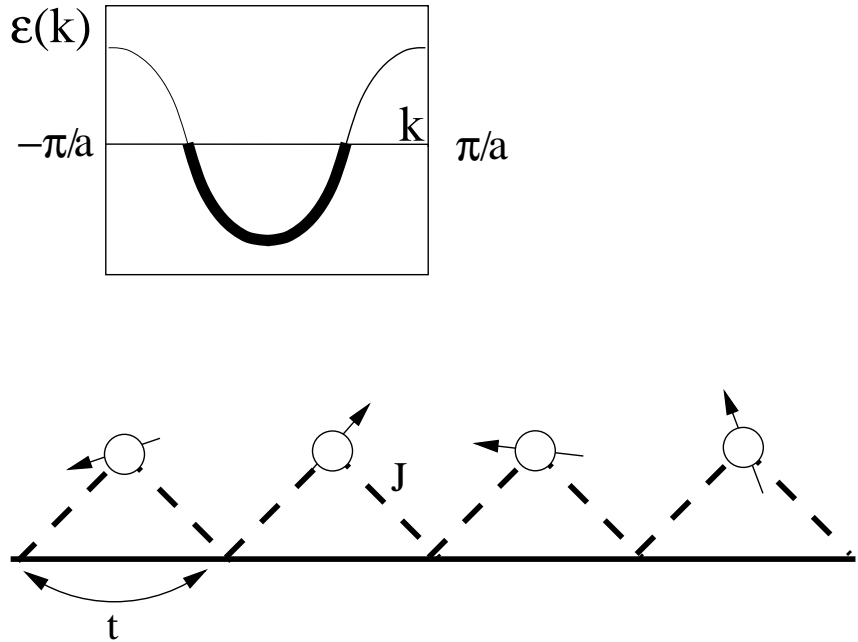}}
\vskip 0.3truein
\protect\caption{
Illustrating the one-dimensional chain which realizes 
a Kondo lattice without back-scattering.}
\label{Fig1}
\end{figure}

The one-dimensional model consists of a tight-binding chain
of conduction electrons. A localized moment is located
between neighboring sites, and couples to them via an antiferromagnetic
Kondo exchange interaction (Fig. \ref{Fig1}.) as follows:
\bea
H = -t \sum _j \biggl[
\psi\dg_{j+1 }
\psi_{j} + {\rm H. c. } \biggr]
+J\sum_j \psi\dg_{j } {\bf \sigma}
\psi_{j }\cdot[ \bf S_{j+\frac{1}{2}}+\bf S_{j-\frac{1}{2}}]\ , 
\eea
where $\psi\dg_j= (\psi\dg_{\up}, \ \psi\dg_{\dw})$ creates the
spin-1/2 conduction electrons, and $\bf S_{j+\frac{1}{2}}$ is a spin-1/2
local moment located between sites $j$ and $j+1$, as shown in Fig. 1.
We begin by linearizing the spectrum around the Fermi energy,
representing the electron on the lattice by a continuum of
right and left moving electrons 
\bea
\frac{1}{\sqrt a}\psi_{j\si} =  \biggl(R_{\si}(x_j)e^{i k_Fx_j}+L_{\si}(x_j)e^{-i k_Fx_j}\biggr),
\eea
to obtain
\bea
H &=& H_0 + V, \cr
H_0 &=& -\ri v_F\int\rd x[R^+_{\s}\p_x R_{\s} - L^+_{\s}\p_x L_{\s}] 
+(\mbox{interaction}),
\label{band}\\
V &=&  v_F \sum_j{\bf S}_j\cdot 
\left[g_f({\bf J}_R + {\bf  J}_L) + g_b\left(\re^{-2\ri k_F x_j}{\bf n}_R + \re^{ 2\ri k_F x_j}{\bf  n}_L\right)\right], \label{one}
\eea
where 
\bea
g_f &=& (J/t),\cr
g_b &=& (J/t) \cos (k_Fa),
\eea
are the dimensionless coupling constants for forward and back-scattering,
$a$ is the lattice spacing and 
\bea
{\bf J}_R =  R^+_{\s}\vec\s_{\s,\s'}R_{\s'}, \: {\bf J}_L =  
L^+_{\s}\vec\s_{\s,\s'}L_{\s'},
\eea
define the currents  of right and left moving electrons.
The back-scattering  term couples the spins to the components of
the staggered magnetization at
momentum $\pm 2k_F$ 
\bea
{\bf n}_R = R^+_{\s}\vec\s_{\s,\s'}L_{\s'}, 
\: {\bf n}_L = L^+_{\s}\vec\s_{\s,\s'}R_{\s'}.
\eea
In general this coupling can not be neglected.
However, if we take the special case of half-filling, where $k_Fa = \pi/2$,
the back-scattering coefficient is identically zero and our
discussion considerably simplifies. 
Note also that we have added an implicit electron-electron interaction
term to $H_0$.  Even though the original mode
contains no explicit interactions,
interactions will be generated by the high-energy physics.  
We shall shortly see how these
implicit interaction effects can  be included into the
bosonized form of the Hamiltonian.

We now focus our
attention on the half-filled case.
Let us begin by reviewing the 
abelian bosonization procedure.
The electron operators
are written
\bea
\left.\begin{array}{c}
R_{\si}(x)\cr
L_{\si}(x)
\end{array}\right\}
= {1 \over \sqrt{2 \pi a }}
e^{-i \sqrt{4\pi} \phi^{\pm}_{\si}(x)  }. \label{fund}
\eea
The right and left-moving electron phases
$\phi^{\pm}_{\si}= \frac{1}{\sqrt{2}}(\phi^{\pm}_c  + \si \phi^{\pm}_s)$
can be written in terms of canonically conjugate fields
\bea
\phi^{\pm}_{\lambda} &= &\frac{1}{2}[\Theta_{\lambda}(x)
\mp
\Phi_{\lambda}(x)], \qquad\qquad (\lambda = c,s)\cr
\Theta_{\lambda}(x) &=& \int^x_{-\infty} dx' \Pi_{\lambda}(x') ,
\eea
where 
\bea
[ \Phi_a(x), \Pi_b(x')] = i\delta(x-x') \delta_{ab} .
\eea
The low-energy physics of the interacting chain 
can then be modelled
by the sum of two Gaussian models for the charge and spin 
fields $\Phi_c$ and $\Phi_s$,
\begin{eqnarray}
H^{(0)} &=& H^{(0)}_c +H^{(0)}_s,\cr
H^{(0)}_c&=&
\frac{v_c}{2}\int \mbox{d}x
\left[K_c \Pi_c^2(x) 
+ K_c^{-1}[\nabla\Phi_c(x)]^2\right],\cr
H^{(0)}_s&=&
\frac{v_s}{2}\int \mbox{d}x
\left[\Pi_s^2(x) + 
[\nabla\Phi_s(x)]^2
\right].
\label{ho}
\end{eqnarray}
Here, the charge and spin velocities $v_{c,s}$
are non-universal and determined by the 
electron interactions in the chain. The spin-stiffness $K_s$
is fixed by the SU(2) spin-rotation symmetry : $K_s = 1$.
$K_c$, the charge stiffness,  sets the charge
susceptibility  of the electron-chain $\chi_c = {\pi K_c /(2 v_c)}$.
$K_c$ is  dependent
on the electron-electron interactions in the chain. 
If these interactions are predominantly  repulsive,
we expect  $K_c<1$.

The bosonized expressions for the spin currents are then 
\bea
J^{(+)}_R &=& J_R^x + i J_R^y = 
\frac{1}{2\pi a_0}\exp[ \ri\sqrt{2\pi}(\Theta_s-\Phi_s)], 
\nonumber\cr
J^{(+)}_{L}& = & J_L^x + i J_L^y =
\frac{1}{2\pi a}\exp[\ri\sqrt{2\pi}(\Theta_s + \Phi_s)],
\nonumber\cr
J_R^z + J_L^z &=& \sqrt{\frac{2}{{\pi}}}\nabla \Phi_s. \label{curr}
\eea
When backscattering is absent,  the charge degrees of freedom  decouple. 
Using  the expressions for 
currents (\ref{curr}) we obtain the Hamiltonian for the 
spin-dynamics
:
\bea
H_{spin} &=& H^{(0)}_s + H_{int},\\
H_{int} &=& v_s
\sum_j \left\{ g^z{\sqrt{\frac {2}{\pi}}} S^z_j\nabla\Phi(j) + 
\frac{g^{\perp}}{2 \pi a}\cos[\sqrt{2\pi}\Phi(j)]
\biggl(
S^+_j
e^{-\ri\sqrt{2\pi} \Theta(j)} + {\rm H. c.} 
\biggr) \right\},
\eea
where we the subscripts $s$ and $f$ on the phase
variables  and coupling constants have been dropped for clarity. 
To examine the model in
a solvable Toulouse limit, an easy-axis anisotropy
has been included into the couplings. 
This  model with   $g^z >> g^{\perp}$ for a single local 
spin was considered by Clarke {\it et al.}\cite{clarke} and is 
equivalent to the two-channel Kondo model.

Following Toulouse, Emery and Kivelson, we absorb the phase factor 
$e^{-\ri\sqrt{2\pi} \Theta(j)}$ into the spin operators by a unitary 
transformation, writing
\bea
\tau^{(\pm)}_j = U\dg S^{(\pm)}_j U=S^{(\pm)}_je^{\mp\ri\sqrt{2\pi} \Theta(j)},
\eea
where
\bea
U = e^{\ri\sqrt{2\pi} \sum_j\Theta(j)S^z_j}.
\eea
Since $[\Theta(j), \Phi(x)] = -i \theta(x_j -x)$, it follows
that
\bea
U \sqrt{2 \pi}\Phi(x) U\dg = \sqrt{2 \pi} \Phi(x) + 2 \pi \sum_l S_l^z 
\theta(x_l-x),
\eea
in other-words, each spin to the right of $x$ 
changes the spin-phase by $\pi$.  This means that 
the electron acquires a phase of $\delta = \pi/2$ each
time it scatters off a spin.  This is the ``resonant scattering''
we expect from the physics of the Kondo effect. 
It also follows that
\bea
\frac{1}{2}\int dx U[\nabla \Phi]^2U\dg = \frac{1}{2}
\int dx[\nabla \Phi]^2 - \sqrt {2 \pi} \sum_j \nabla \Phi(x_j) S_j^z\ ,
\eea
so that 
\bea
H^* &= &U H U\dg,  \cr
&=&  
H^{(0)}_s
+ v_s\sum_j \biggl\{
(g^z-\pi){\sqrt{\frac {2}{\pi}}} \tau^z_j\nabla\Phi(j) + 
(-1)^j\frac{g^{\perp}}{ \pi a}\cos[\sqrt{2\pi}\Phi(j)]
\tau^x_j
\biggr\}.
\eea
To characterize
the low energy physics of the ZKE model, it is convenient
to examine the strongly anisotropic ``Toulouse limit''  where
$g_z= \pi$, so 
\bea
H^* =  
H^{(0)}_s
+v_s\sum_j \frac{g^{\perp}}{ \pi a}(-1)^j\tau^x_j
\cos[\sqrt{2\pi}\Phi_s(j)].
\eea
Experience gained from the one and 
and two-channel Kondo models leads us to  anticipate that provided
the local moments are screened, then the 
physics of the Toulouse limit will extend 
extending out to the isotropic point.

\section{The order parameter}

In this section we discuss  the correlations present
in the ground-state of the  ZKE model at the Toulouse
point.  This model is, in essence, 
a chain of two-channel Kondo impurities: each localized spin is
coupled to the right and left-moving screening channels.  In isolation, 
a two-channel Kondo impurity retains an unquenched degree of freedom
associated with the ability of the Kondo singlet to fluctuate between the
two screening channels.  This residual spinless degree of freedom 
behaves like a localized Majorana fermion.  In the ZKE model, 
these degrees of freedom become coupled, removing the 
residual entropy by generating a low-lying band of spinless excitations. 

At the Toulouse point, the $\tau^x_j$ commute with the Hamiltonian,
becoming constants of the motion with eigenvalues $\tau^x_j = \pm \frac{1}{2}$.
In the ground-state, the spin-phase of the conduction chain prefers
to acquire a constant value.  The $(-1)^j$ coupling term in the Hamiltonian
will mean that the $\tau_j^x$ develop staggered long-range order in the
ground-state
\bea
\langle \tau^x_j \rangle =\frac {Z}{2} (-1)^j.\label{staggered}
\eea
The effective spin Hamiltonian for the ground-state is then a 
sine-Gordon model
\bea
\tilde H =  
H^{(0)}_s
+\frac{M_o}{ \pi a}\int dx \cos[\sqrt{2\pi}\Phi_s(x)], \label{gordon}
\eea
where $M_o= \Lambda g^{\perp}$ is the ``bare'' mass and $\Lambda = v_s/a$
is the high-energy cut-off. 
The ``$\sqrt {2 \pi}$''
prefactor in the cosine guarantees
that the sine-Gordon Hamiltonian (\ref{gordon}) 
possesses 
a full SU(2)-spin symmetry (see \cite{haldane}). In other-words, the formation of 
Kondo singlets completely quenches the anisotropy of the Kondo coupling,
restoring the full symmetry of the band. 
From the Bethe Ansatz solution to the
sine Gordon model, it is
known that the spectrum of this model contains 
a low-lying triplet separated from the 
ground-state by gap  
$\Delta_s$ and a singlet with a gap $M = \sqrt 3 \Delta_s$ \cite{affleck}. 
The approximate size of the spin-gap  $\Delta_s$ can be obtained from
simple scaling arguments. Since the spin phase $\Phi_s$
is a Gaussian variable, we know that in the uncoupled chain,
$\cos(\sqrt{2\pi}\Phi)$ has a power-law correlations
\bea
\langle \cos(\sqrt{2\pi}\Phi(1))\cos(\sqrt{2\pi}\Phi(2))\rangle
\sim \frac{1}
{(\Delta x^2 - v_s^2 \Delta t^2)^
{\frac{1}{2}}
},
\eea
corresponding to scaling dimension $d=1/2$.  When we
integrate out high frequency modes, 
rescaling $(x,t) = \lambda (x',t')$, 
where $\lambda = (\Lambda/\Lambda')$ is the ratio of cut-off energies,
we must
rescale the operator
\bea
\cos(\sqrt{2\pi}\Phi)= \lambda^{-\frac{1}{2}} \cos(\sqrt{2\pi}\Phi').
\eea
The coupling  constant then scales 
$
g_{\perp} \rarrow \lambda^{(2-\frac{1}{2})}g_{\perp} = g_{\perp}^*
$.  
The spin gap develops at the point when strong-coupling is reached.
Setting $g_{\perp}^*=1$, $\lambda = \frac{\Lambda}{\Delta_s}$, where $\Lambda =
v_s/a$ is the upper cut-off, it follows that 
$\Delta_s \sim \Lambda (g^{\perp})^{2/3}$. 
This gap is much greater than the
single impurity Kondo temperature $T_K \sim \Lambda (g^{\perp})^{2}$.

Although there is no single operator that can be directly related
to the variable $\tau^x_j$, there is a collection of composite operators
that are equal to $\tau^x_j$, up to a phase factor. 
Consider the following composite operators
\def\jpl{j+\frac{1}{2}}
\def\jmn{j-\frac{1}{2}}
\bea
\Psi^{+}_j &= &-{i}\biggl(\psi\dg_{\jpl}\vec \si \si_2\psi\dg_{\jmn}
\biggr)\cdot \vec S_j,\cr
\Psi^{z}_j &= &\frac{(-1)^j}{2}
\biggl(\psi\dg_{\jpl}\vec \si \psi_{\jmn}
+\psi\dg_{\jmn}\vec \si \psi\dg_{\jpl}
\biggr)\cdot \vec S_j. \label{oddfr}
\eea
The first operator describes a composite singlet formed between
a local moment and a triplet pair on the neigboring
sites, the second describes a singlet
between the local moment and an
electron delocalized on the two neighboring sites. 
These operators as the order parameters
for odd-frequency singlet pairing and odd-frequency charge-density
wave formation respectively. 
An expectation value $\langle\Psi^+_j\rangle$ would break 
gauge invariance, but it does not
induce any equal-time pairing. 
$\Psi$ changes sign under an exchange
of electron spin or position co-ordinates. 
The 
induced pair correlation function 
\bea
F_{\alpha \beta}(x-x',t-t') = \langle \psi_{\alpha}(x,t)\psi_{\beta}(x',t')
\rangle.
\eea
must exhibit the same
symmetries, i.e. 
\bea
F_{\alpha \beta}(x,t) = S F_{\beta \alpha}(x,t) = P
F_{\alpha \beta}(-x,t), \qquad
P=S=-1.
\eea
Since 
the parity for the 
combined operation of spin, space and time inversion is
$SPT=-1$ and $SP=1$, it follows that $T=-1$, i.e
the pair correlations 
are {\sl odd} in time 
\bea
F_{\alpha \beta}(x,t) = - F_{\alpha \beta}(x,-t).\eea
$\Psi^z_j$  is obtained by taking
the commutator of $\Psi_j^+$ with the staggered isospin
operator ${\cal T}^- = \sum (-1)^j \psi_{j\down}\psi_{j\up}$,
$\Psi^z_j = [ {\cal T}^- , \Psi^+_j]$. Since the
action of ${\cal T}^-$ is to convert a singlet pairing field
to a charge-density operator, it follows that 
an expectation value $\langle \Psi^z_j\rangle $ induces an 
odd-frequency charge modulation. 

The long-wavelength decompositions of the composite order parameters
are 
\bea
\Psi^{+}_j &\sim &a (-1)^j \left[R\dg_{\up}(x_j)L\dg_{\up}(x_j)S^-_j
+ R\dg_{\down}(x_j)L\dg_{\down}(x_j)S^+_j\right]+ \dots\ , \cr
\Psi^{z}_j &\sim & (a (-1)^j/2)
\left\{
 \left[R\dg_{\up}(x_j)L_{\down}(x_j)-
 L\dg_{\up}(x_j)R_{\up}(x_j)\right] S^-_j- {\rm H. c} \right\}+ \dots \ , 
\label{replace}
\eea
where the terms coupling to $S_z$ have been omitted. 
We may re-write the operators appearing in these expressions
using the bosonized expressions for the Fermi fields(\ref{fund}), 
\bea
(-1)^jR^+_{\up}(x_j)L^+_{\up}(x_j)S^-_j &=& 
 {i \over 2 \pi a }(-1)^j\tau^-_j
\exp[-\ri\sqrt{2\pi}\Theta_c],\nonumber\cr
(-1)^jR^+_{\down}(x_j)L^+_{\down}(x_j)S^+_j 
&=&{i \over 2 \pi a }(-1)^j\tau^+_j
\exp[-\ri\sqrt{2\pi}\Theta_c],\nonumber
\cr
(-1)^jR^+_{\up}(x_j)L_{\down}(x_j)S^-_j 
&=& {1 \over 2 \pi a }(-1)^j\tau^-_j
\exp[-\ri\sqrt{2\pi}\Phi_c]\nonumber,\cr
(-1)^jR^+_{\down}(x_j)L_{\up}(x_j)S^+_j 
&=& {1 \over 2 \pi a }(-1)^j\tau^-_j
\exp[-\ri\sqrt{2\pi}\Phi_c],\nonumber
\eea
where 
$\tau_j^{\pm} = \exp [\mp \ri \sqrt{2 \pi} \Theta_s] S^{\pm}_j$.
Using (\ref{staggered}), we obtain
\bea
\Psi^+_j  &\sim &\frac{Z}{\pi} e^{-i \sqrt{2 \pi}  \Theta_c(j)},\nonumber\cr
\Psi^z_j  &\sim &\frac{Z}{\pi} \sin{ \sqrt{2 \pi}  \Phi_c(j)}.
\eea
The scaling dimensions of $e^{-i\sqrt{2 \pi} \Theta_c}$
and $e^{-i\sqrt{2 \pi} \Phi_c}$ are $\frac{1}{2K_c}$ and
$\frac{K_c}{2}$ respectively, so that
\bea
\langle \Psi^-(1)\Psi^+(2)\rangle &\sim& \frac{1}{[ \Delta x^2 - v_s^2 
\Delta t ^2 ] ^{\frac{1}{2K_c}}}\nonumber,\cr
\langle \Psi^z(1)\Psi^z(2)\rangle &\sim& \frac{1}{[ \Delta x^2 - v_s^2 
\Delta t ^2 ] ^{\frac{K_c}{2}}}.
\eea
In other words, the development of long-range order
in the variable $\tau_j^x$ leads to long-range 
odd-frequency singlet  and odd-frequency charge-density wave correlations,
where $\Theta_c$ is the phase of the pair correlations and $\Phi_c$
is the phase of the charge-density wave correlations. 
Odd-frequency pair correlations will dominate the long-range
correlations when the electron interactions are repulsive and $K_c<1$.

Of course, since the Toulouse limit  is anisotropic,  a certain
amount of singlet pairing corrleations are induced, for example,
the triplet order parameter
\bea
\phi^t_j = \biggl[\psi\dg_{\jpl  \up} \psi\dg_{\jmn\down}
-\psi\dg_{\jmn  \up} \psi\dg_{\jpl\down}\biggr]
\sim  \frac{1 }{2 \pi} \langle\cos \sqrt{2 \pi}  \Phi_s(j)\rangle
\exp[ -i \sqrt{2 \pi} \Theta_c(j)] \eea
also develops long-range correlations. 
Since the scaling dimension of $\cos \sqrt{2 \pi}  \Phi_s$ is 
$d=\frac{1}{2}$, we expect 
$\vert \langle 
\cos \sqrt{2 \pi}  \Phi_s(j) \rangle\vert \sim (\Delta_s/t)^{\frac{1}{2}} \sim 
(J/t)^{\frac{1}{3}}$, 
so the amplitude of triplet pair correlations is reduced
relative to the odd-frequency singlet correlations by a
factor
$(J/t)^{\frac{2}{3}}<<1$. These secondary triplet
 correlations are  induced by the anisotropy, 
and will vanish in the isotropic limit. 

We thus see that in the absence of back-scattering, 
the main characteristic of the ZKE model is the development of
a spin gap and, in the case of repulsive electron-electron
interactions,  the establishment of long-range
odd-frequency singlet pair correlations in its ground-state.

\section{Excitations}

Let us discuss the excitation spectrum  of the Hamiltonian 
(\ref{gordon}) in more detail. 
We assume that the spin field is locked, so that  
$\cos(\sqrt{2\pi}\Phi_s)$ can be replaced by its average. 
This yields  
the following effective ``magnetic'' field acting on $\tau^x$:
\begin{equation}
h = \Lambda (g^{\perp})\la\cos(\sqrt{2\pi}\Phi_s)\ra \sim 
\Lambda (g^{\perp}/a_0)^{4/3} \sim  (\Delta_s)^2/\Lambda .
\end{equation}
Thus the energy necessary to flip a single pseudospin is much smaller than 
the gap of the propagating spin excitations. Exactly at the Toulouse limit 
pseudospin flips  do not propagate, but this changes when one considers 
finite $\delta g^z = \pi - g^z$.  
We can integrate approximately  over $\Phi$ putting  
\begin{equation}
\la\la\nabla\Phi_s(q, \omega)\nabla_x\Phi_s(-q , -\omega)\ra
 \sim  \chi_s
\frac{(q)^2}{q^2+ \xi^{-2}},
\end{equation}
where $\chi_s\sim 1 / \Lambda$  and $\xi = a \Lambda/\Delta_s$ is the correlation
length. 
This leads to the following quantum Ising model
Hamiltonian for the coupled pseudospins:
\begin{equation}
H = \sum_j[ \tilde J(q) \tau^z(q)\tau^z(-q)] +h\sum _j(-1)^j\tau_j^x,
\end{equation}
where $\tilde J(q) = \chi(q) [\Lambda \delta g]^2$.  We can get a
rough idea of the dispersion of the pseudo-spin excitations using a
Holstein-Primakov transformation $\tau^+_q = b\dg _q$, which gives a
spectrum 
\bea \omega_q = \biggl[ h(\tilde J(q) +
h)\biggr]^{\frac{1}{2}}.  \eea 
We see that in a narrow range of
wavevectors $|q| < \xi^{-1} << a^{-1}$ where $a$ is the lattice constant,
$\omega_q \sim h$.  Outside this region, $\omega_q \sim \Delta_s
\bigl[(\delta g^z)^2 + (\Delta_s/\Lambda)^2 ]^{\frac{1}{2}}$.  There are thus
two gaps in the excitation spectrum:
\begin{itemize}

\item Spin-gap $\Delta_s = \Lambda  (g^{\perp})^{\frac{2}{3}}$.

\item Pseudo-spin gap $h = (\Delta_s)^2 /\Lambda$.

\end{itemize}
The lower band of spinless, dispersing excitations is most
naturally interpreted as the 
residue of the Majorana excitations present in individual two-channel Kondo
impurities. 

Let us now develop a 
heuristic picture of the temperature dependence of
the ZKE chain. At the highest possible temperatures,
the individual Kondo spins are unbound, with a spin susceptibilty
$\chi_0\sim 1/T$. A ``Zhang-Rice'' singlet\cite{zhang} will begin to form
around each spin along the chain at a characteristic scale $T^*$.
We may estimate this scale from a 
high temperature expansion. 
At high temperatures $T >> h$, 
\begin{equation}
\Pi \equiv \la\la\cos(\sqrt{2\pi}\Phi)\cos(\sqrt{2\pi}\Phi)
\ra\ra_{\omega,q \rightarrow 0} 
\sim 1/T,
\end{equation}
so the RPA expression for the pseudo-spin  susceptibility 
\bea
\la\tau^x(-\omega, -q)\tau^x(\omega, q)\ra
_{\omega,q=0} 
= [(\chi_0)^{-1} - (g^{\perp})^{2}\Pi]^{-1}\sim 
(T - \mbox{const}(g^{\perp})^{2}/T)^{-1}
\eea
acquires a singularity at $T^* \sim \Lambda g^{\perp}$. 
This singularity will be smeared out by fluctuations, but
marks the development of Zhang-Rice singlets between the local
moments and the conduction chain. 
For small $g^{\perp}$,  $T^*$ is much smaller than the
zero-temperature spin-gap, but much larger than the
pseud-spin gap $h$, $h<< T^*<<\Delta_s$.
There is thus a 
wide temperature region,  
$h(0) << T << T^*$, 
where  the pseudo-spin band is nondegenerate, but the 
local spins are strongly correlated with the conduction electrons 
to form Zhang-Rice singlets. 

\section{Re-introduction of back-scattering.}

We now return to discuss the back-scattering
terms. We should like to be sure that our results
are indeed robust against the inclusion of small amounts of
back-scattering. 
At half-filling,
we can not really turn on the interactions between the
electrons in the chain, 
for in this case the model  will become develop a charge gap. 
But we need to turn on the electron-electron
interactions, because only then will the odd-frequency pair
correlations become enhanced over the odd-frequency charge
correlations.   The key to this dilemma, is to dope the model
away from half-filling. This introduces small amounts
of back-scattering. Writing
\bea
n^3_{R,L} &=& -\frac{1}{2 \pi a} \sin (\sqrt{2 \pi} \Phi_s) e^{\mp i \sqrt{2 \pi}\Phi_c},\cr
n_R^{\pm}&=&  -\frac{1}{2 \pi a}e^{-i \sqrt{2 \pi}(\Phi_c\mp \Theta_s)},\cr
n_L^{\pm}&=&  -\frac{1}{2 \pi a}e^{-i \sqrt{2 \pi}(-\Phi_c\mp \Theta_s)},
\eea
the back-scattering part of the Hamiltonian may be written
form 
\bea
H_{int} =
 \sum_j\left\{\cos(\sqrt{2\pi}\Phi_c + 2k_FR_j)\left[g_b^z\tau^x_j + g_b^{\perp}(-1)^j\tau^z_j\sin[\sqrt{2\pi}\Phi_s(j)]\right]\right\}.
\eea
Although this term is oscillatory in nature, we need to 
examine its scaling properties to ensure that 
incommensurate phases do not form before 
odd-frequency pairing has time to develop. 
The scaling dimension of the back-scattering term is $K_c/2$ 
(the first term has a larger scaling dimension $(K_c+1)/2$, and
can be neglected), so the coupling constant rescales to a renormalized
value 
\bea
g^*_b[\omega]\sim g_b\biggl(\frac {\omega}{\Lambda}
\biggr)^{\frac{K_c}{2}-2}.
\eea
The forward scattering scales to strong coupling at 
at energies $\omega$
comparable with the spin gap $\Delta_s$. In order that odd-frequency
correlations develop, we require
that the renormalized backscattering coupling 
constant is small at this scale, i.e. 
\bea
g^*_b[\Delta_s] \sim g_b\biggl(\frac {\Delta_s}{\Lambda}
\biggr)^{\frac{K_c}{2}-2} \sim 
g_b\biggl(g_f^{\perp}\biggr)^{(K_c-4)/3} << 1, \label{ink}
\eea   
where we have used $\Delta_s = \Lambda g_f^{2/3}$.
But $g_b/g_f= \cos (k_F a)$, so that 
\bea
\cos k_Fa << \biggl(\frac{J}{t}\biggr)^{(1-K_c)/3},
\eea
defines the region around half-filling where we expect odd-frequency
pair correlations to survive in the original model.  
For repulsive interactions $1/2<K_c<1$,
so that even in the limit of infinitely
strong repulsion, $K_c=1/2$, this  condition allows for
a broad range of doping.  For these reasons, we expect odd-frequency
pairing correlations to persist in a finite region  around half-filling. 

\section{Discussion: into Three dimensions.}

\begin{figure}[tb]
% ********   This is for two columns
\epsfxsize=3.0in 
% ***********For one column  ********************
%\epsfysize=5.5in 
% ***********************************8
\centerline{\epsfbox{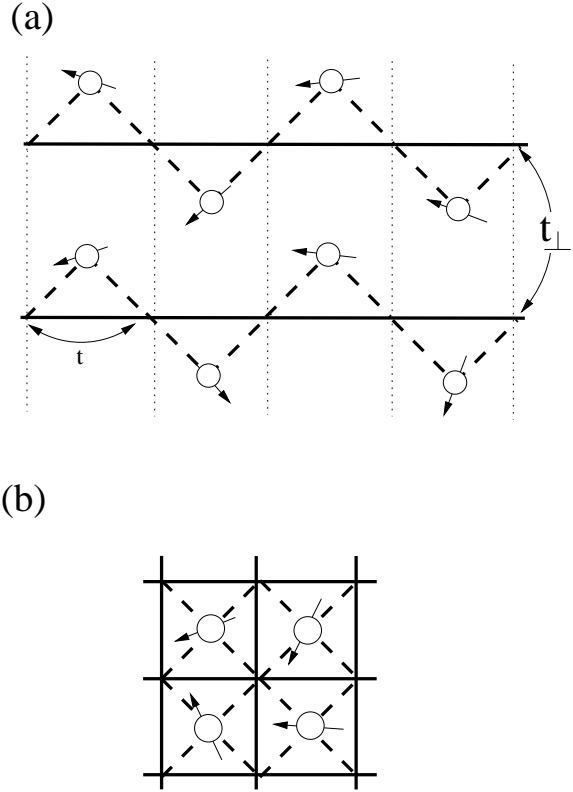}}
\protect\caption{
(a) Coupled ZKE chains. 
(b)``Confederate Flag'' model, a symmetric 
generalization of the ZKE model to two dimensions.
Unlike the ``Zhang-Rice'' singlet, where each spin couples
to a single Wannier state built up out of four orbitals, here
the back-scattering is absent, so that each 
spin has {\sl four} separate antiferromagnetic links to neighboring
electrons. 
}
\label{Fig2}
\end{figure}

We should like to end this paper by discussing how the results of
the ZKE model might extend to higher dimensional models.
To assemble the chains into a three dimensional structure 
one may introduce a direct electron hopping $t_{\perp}$ between the chains,
as shown in Fig. \ref{Fig2}(a)
We consider the case where this hopping is small: $|t_{\perp}| << \Delta_s$,  
so that single particle hopping is 
virtual, but pair hopping is direct. There will be Josephson tunneling 
of both ordinary and composite (\ref{oddfr}) pairs, but with very different 
matrix elements. For ordinary pairs the hopping matrix element is 
$\sim t_{\perp}^2$, 
but for a composite pair in the absence of {\it direct} spin 
exchange between the chains it is $\sim t_{\perp}^4$. 
Taking into account Eq.(\ref{replace}) we get the same effective 
interaction in both cases:
\begin{equation}
V_{int} = \sum_{i \neq j}T_{ij}\tau^x_i\tau^x_j\cos[\sqrt{2\pi}(\Theta_i - \Theta_j)].
\end{equation}
This form of 
interaction was postulated by Abrahams {\it et al.} in 
their last paper about odd-frequency pairing \cite{abrahams}.

If the composite order parameter has the long-range correlations
in the individual chains, then the bare susceptibility for composite
pairing in the system of uncoupled chains has the following 
frequency dependence
\bea
\chi^{(o)}(\omega) \sim \int dx dt e^{i \omega t}(x^2 - v_c^2 t^2)
^{-\frac{1}{2K_c}} \sim \omega^{-2+(1/K_c)}.
\eea
At a temperature $T$, the composite pair susceptibility
\bea
\chi^{(o)}(T)\sim T^{-2+({1}/{K_c})}
\eea
is a divergent function of temperature, providing $K_c > \frac{1}{2}$, 
a condition satisfied except in the extreme limit of repulsive
interactions.  For weak interchain coupling, the effective pair
susceptibility $\chi(T) = [(\chi^{(o)})^{-1}- \langle T_{ij}\rangle ]^{-1}$
will diverge at a temperature $T_c \sim \langle T_{ij}\rangle ^{\eta}$ with $\eta = K_c/(2K_c - 1)$, giving rise to a 
a macroscopically phase coherent odd-frequency superconductor. 

How can we study the odd-frequency  pair condensate which forms?
One proposal, is to examine the limit of strong intrchain repulsion, for 
in this limit the superconducting order 
parameter has scaling dimension $(1/K_c)\sim 2$, so it
can be represented as the fermion bilinear. In this case
one can describe 
the low-energy behaviour by 3D fermionic Hamiltonian of spinless fermions
\bea
H_{eff}& =& \sum_j H(j) + \sum_{i,j}T_{ij}(R^+_iL^+_iR_jL_j + H.c)],\cr
H(j)&=& v_c\int \rd x[ - \ri(R_j^+\p_xR_j - L^+\p_x L) - g_0R_j^+R_jL^+_jL_j ],
\eea
where $g_0 \sim \frac{1}{2} - K_c$. 
This Hamiltonian, which  describes a  hopping of weakly coupled 
preformed pairs, bears remarkable resemblence to 
the Hamiltonian introduced by Anderson and Chakravarthy \cite{pwa} to describe the formation of the SC order parameter in 
cuprates. The perturbative study of this Hamiltonian may provide insights
into the properties of the coupled chain system.

Another approach which seems promising, is to use a slave-fermion representation 
for the localized spins:
\bea
\vec S_j = f_{j\alpha}\dg\biggl(\frac{\vec \sigma_{\alpha \beta}}{2}
\biggr) f_{j\beta},
\eea
together with the constraint $n_f(j)=1$ This representation  has
a local $SU(2)$ symmetry\cite{affleck2}.  By carrying out
a Hubbard Stratonovich decoupling of the Kondo interaction which
respects this gauge symmetry, one may transform the Kondo interaction 
into the form\cite{natan}
\bea
J\sum_{\lambda=1,p}\vec \si_{\lambda}\cdot \vec S_j
\longrightarrow
\sum_{\lambda=1,p}\left\{\tilde  c_{\lambda} \dg {\cal V}^{\lambda\dagger}(j) \tilde f_j
+\tilde  f_j \dg {\cal V}^{\lambda}(j) \tilde c_{\lambda}
 + \frac{1}{2J} Tr[ {{\cal V}}^{\lambda\dagger}(j){\cal V}^{\lambda}(j)]\right\},
\eea
where
\bea 
\tilde  f_j =
\left(\matrix{f _{ j\uparrow}\cr 
f\dg _{j\downarrow}}\right), \qquad
\tilde { c}_{\lambda } =
\left(\matrix{c _{\lambda \uparrow}\cr 
c\dg _{\lambda \downarrow}}\right), 
\eea
are the Nambu spinors for the slave fermion and the 
conduction electrons.  
${{\cal V}}^{\lambda }(j)= 
i {\rm V^{\lambda}}(j) e^{i \vec \theta_{\lambda}(j)n_{\lambda}(j) \cdot \vec \tau}$ is 
an SU(2) matrix representing the singlet bond formed
between the spin at site $j$ and its neighbor at site $\lambda$. The quantity $p$ is the number
of orbitals that hybridize with the local moment. For the ZKE model,
$p=2$.
This Hamiltonian has the $SU(2)$ gauge symmetry
\bea
  {\cal V}^{\lambda}(j) & \rightarrow&
 g_j { {\cal V}}^{\lambda}(j), \cr
\tilde { f}_j &\rightarrow& g_j \tilde { f}_j,
\eea
where $g_j$ is an $SU(2)$ operator.
Although ${ {{\cal V}}}^{\lambda}$  is not gauge invariant, if there
is more than one neighbor, then ${\cal V}^{\lambda\dagger}{\cal V}^{\lambda'}$
is an SU(2) invariant which describes the phase coherence
of the Kondo singlet between the neighboring atoms.  
For the ZKE model, this invariant
is directly related to the two composite order parameters
\bea
{\cal V}^{j+1\dagger}{\cal V}^{j}\propto
\left[ \matrix{
(-1)^j\Psi^z_j & \Psi^{-}_j \cr
\Psi^{+}_j& (-1)^{(j+1)}\Psi^z_j }
\right].
\eea
This relation expresses the basic result that 
phase-coherence between a Kondo singlets
that is distributed over more than one site, gives rise
to odd-frequency correlations. 
This type of mean-field theory can  be
tested by checking that the mean-field theory with 
Gaussian fluctuations is able to reproduces the sentient
feature of the ZKE bosonization.  Its  virtue of course lies
in its ability to be generalized to a  higher dimensional
model. One particularly interesting model in this respect
is the ``confederate flag model'' shown in Fig. \ref{Fig2}(b).
This model is reminiscent of the s-d models
used for cuprate superconductors\cite{zhang,maekawa}, but 
the Kondo scattering  in each plaquet
has been artificially stripped of  the back-scattering
terms which simultaneously hop an electron and flip a
spin, to create an interaction
\bea
H_{int} = J \vec S_j \cdot [ 	\vec \sigma_1 +\vec \sigma_2 +\vec \sigma_3+\vec \sigma_4]
\eea
at each plaquet. It will be very interesting to see
whether the removal of back-scattering does indeed give rise
to coherent odd-frequency singlet pairing. 

In this paper we have discussed recent  non-perturbative 
results due to Zachar, Kivelson and Emery\cite{zachar,zachar2} which suggest that
odd-frequency pair correlations  develop in
a one-dimensional Kondo lattice where back-scattering is suppressed.
We have introduced a simple
lattice model where back-scattering is naturally suppressed,
and argued that
that it is odd-frequency singlet pairing that develops in this
model. These correlations are robust
against doping around half-filling. We have argued that when
ZKE chains are coupled, this will lead to true long-range
odd-frequency singlet pairing.  Finally, we have proposed
that  the SU(2) approach to the Kondo lattice model offers
a natural way to study this phenomenon in higher dimensional models.

\section{Acknowledgements}
We are grateful to S. Kivelson, A. Schofield   and D. L. Cox for valuable   
discussions and illuminating remarks.
This research was supported in part by the National Science 
Foundation under Grants No. PHY94-07194
and DMR-93-12138. P. C. and A. M. T. acknowledge
a support from  NATO under Research Grant CRG. 940040.

\end{document}